# Zero-bias photocurrents in highly-disordered networks of Ge and Si nanowires

M. Golam Rabbani[1], Sunil R. Patil[1,2], Amit Verma[3], Julian E. Villarreal[4], Brian A. Korgel[5], Reza Nekovei[3], Mahmoud M. Khader[5], R. B. Darling[1], M. P. Anantram[1]

[1]Department of Electrical Engineering, University of Washington, Seattle, WA 98195 USA,

[2]Department of Physics, College of Engineering, Pune, 411005 MS, India,

[3]Department of Electrical Engineering and Computer Science, Texas A&M University – Kingsville, Kingsville, Texas 78363, USA,

[4]McKetta Department of Chemical Engineering, Texas Materials Institute, Center for Nano- and Molecular Science and Technology, The University of Texas at Austin, Austin, Texas 78712, USA,

[5]Gas Processing Center, College of Engineering, Qatar University, Doha, P.O. 2713, Qatar.

## Abstract

Semiconducting nanowire (NW) devices have garnered attention in self-powered electronic and optoelectronic applications. This work explores and exhibits, for the first time for visible light, a clear evidence of the zero-biased optoelectronic switching in randomly dispersed Ge and Si NW networks. The test bench, on which the NWs were dispersed for optoelectronic characterization, was fabricated using standard CMOS fabrication process, and utilized metal contacts with dissimilar work functions - Al and Ni. The randomly dispersed NWs respond to light by exhibiting substantial photocurrents and, most remarkably, demonstrate zero-bias photo-switching. The magnitude of the photocurrent is dependent on the NW material, as well as the channel length. The photocurrent in randomly dispersed GeNWs was found to be higher by orders of magnitude compared to SiNWs. In both of these material systems, when the length of the NWs was comparable to the channel length, the currents in sparse NW networks were found to be higher than that in dense NW networks, which can be explained by considering various possible arrangements of NWs in these devices.

I. **Introduction:**

Semiconducting nanowires, which offer long absorption paths along their length yet shorter distances for carrier collection/transport[1,2], have been widely studied for prototype applications, such as solar cells[3], photo-detectors[4] and nanolasers[5]. Often, these studies encompass single NWs and involve complex and costly nano-fabrication and ultra-sensitive measurements to achieve unique applications such as single-photon optics[6], self-powered photo detection[4], etc. However, given the fabrication challenges with single NWs, the potential of NW-based devices comprised of NW mats, networks or randomly dispersed NWs (RDNWs)[7] needs to be understood and exploited. The NW networks, mats and RDNWs are relatively cheaper and simple to fabricate and also have important scientific and technological advantages over single NWs, such as effective light trapping and refractive-index matching[7,8]. Chemical sensors fabricated from NW mats are more sensitive than that of single NW devices[9]. Other advantages of networks and mats are that they are relatively strong and flexible, self-supported, and thin (approximately 50 μm) with ~90% void space, which can absorb most of the light from the ultraviolet (UV) to near-infrared (IR) wavelength ranges[10].

An important aspect of nanoscale materials, such as NWs, is their behavior at interfaces, which often overrides the active nano-material physics and can potentially alter the overall performance of the electronic and optoelectronic devices[11]. In optoelectronic applications, the interfaces are generally the electronic contacts to the NWs, which collect photogenerated electrons and holes upon illumination of semiconductor NWs. The interfaces and their influences on the overall device performance have been studied in electronic devices, even at the nanometer scale[12,13], and observations have shown that nano-contact resistance can dominate the overall electrical properties.

Apart from collection of charge carriers, the idea that work-function asymmetry in contacts can induce an electric field in the active channel—which in turn directs carriers—has attracted the attention of the optoelectronic research community. Note that such an asymmetry in bulk semiconductors is achieved by chemical doping, which becomes extremely difficult to control at the nanoscales. A recent modeling study of dual-metal Schottky contacts for nano and micro-wire solar cells[14] reported that dissimilar metal contacts can be an excellent alternative to dissimilar doping, which in turn enhances the fabrication reliability, particularly where nanoscale materials are involved. The experimental investigations of asymmetric contacts for photo-switching behavior have been carried out in single NW devices[15–17]. However, for nanowire networks, observation of photo-switching has only been limited to ZnO NWs that too only for UV excitations[18]. This effect is largely due to the inherent losses originating from misalignments and inter-NW transport, which is not clearly understood.

Over the last two decades, NW synthesis has advanced considerably. Vapor-liquid-solid (VLS) growth is predominantly used for synthesis of one-dimensional structures of all kinds of materials[19–23]. Another method is the so-called supercritical fluid-liquid-solid (SFLS) approach[24,25]. The VLS method is slow and expensive, whereas SFLS has better control over NW size (i.e., diameter heterogeneity) and allows for the synthesis of large (industrially relevant) quantity of NWs[24]. NWs synthesized in the SFLS approach are single-crystalline[24,25] and can be free of metal particles[26]. It is also possible to synthesize NWs of compound materials[25], and surface modifications[27,28] can be done simultaneously with synthesis. However, NWs synthesized by the SFLS route have not been extensively explored for device applications as have NWs synthesized through VLS.

In this work, we report the detailed experimental characterization of the optoelectronic switching response of highly disordered networks of SFLS-synthesized Ge and Si NWs between asymmetric work-function contacts under red light excitation. The effect of density of the networks and distance between contacts on the performance is also investigated. The rest of the report is organized as follows: Section II presents details of NW synthesis and RDNW-based device fabrications, section III demonstrates and discusses the results of photo-response in Ge and Si RDNWs, followed by the section IV that summarizes and concludes the work.

## II. Experimental
### a. Synthesis of GeNWs and SiNWs

Ge and Si NWs were synthesized via metal nanocrystal-seeded SFLS method using diphenylgermane (DPG) and monophenylsilane (MPS), respectively, as the precursors. To synthesize Au-seeded GeNWs (diameter=45±15nm and few to several tens of microns long), the reactant solution was prepared using 16 mg/L of Au nanocrystals and 34 mM DPG (Gelest, > 95%) in anhydrous toluene. The synthesis reaction was carried out at 380°C, 6.2 MPa and at a flow rate of 0.5 mL/min for 40 min. Subsequently, the reactor was allowed to cool isochorically to 80°C for surface passivation via thiogermylation[29]. Inside a glovebox, a solution was prepared containing 8 mL of anhydrous toluene and 4 mL of 1-dodecanethiol (Sigma-Aldrich, ≥98%) as a passivating agent. The passivation solution was injected into the reactor at a flow rate of 2 mL/min, until the pressure returned to 6.2 MPa (the exit valve of the reactor remained closed during injection). The reactor was then sealed and kept at 80°C for 2 h. The NW products were removed and were washed three times by dispersion in a solution of toluene, hexanes, and chloroform (2:1:1) and precipitated via centrifugation between each washing step.

The Sn-seeded SiNWs were synthesized with MPS according to a procedure described by Bogart et al.[30]. In a typical synthesis, a reactant solution containing 30 mL of anhydrous toluene, 500 μL of MPS and 96 μL of bis(bis(trimethylsilyl)amino)tin (Sn(HMDS)$_2$, Sigma-Aldrich, 99.8%) was prepared inside an argon-filled glovebox and placed inside a stainless steel injection cylinder. This solution was fed to a titanium reactor filled with anhydrous toluene at 490°C and 10.3 MPa, at a rate of 0.5 mL/min for 40 min. Afterwards, the reactor was cooled, the contents extracted, and the product washed as in the case of the GeNWs. Physical characterizations of NWs synthesized in the above processes have been reported in detail elsewhere[27,29,30].

### b. Fabrication of dual-metal test benches on oxide

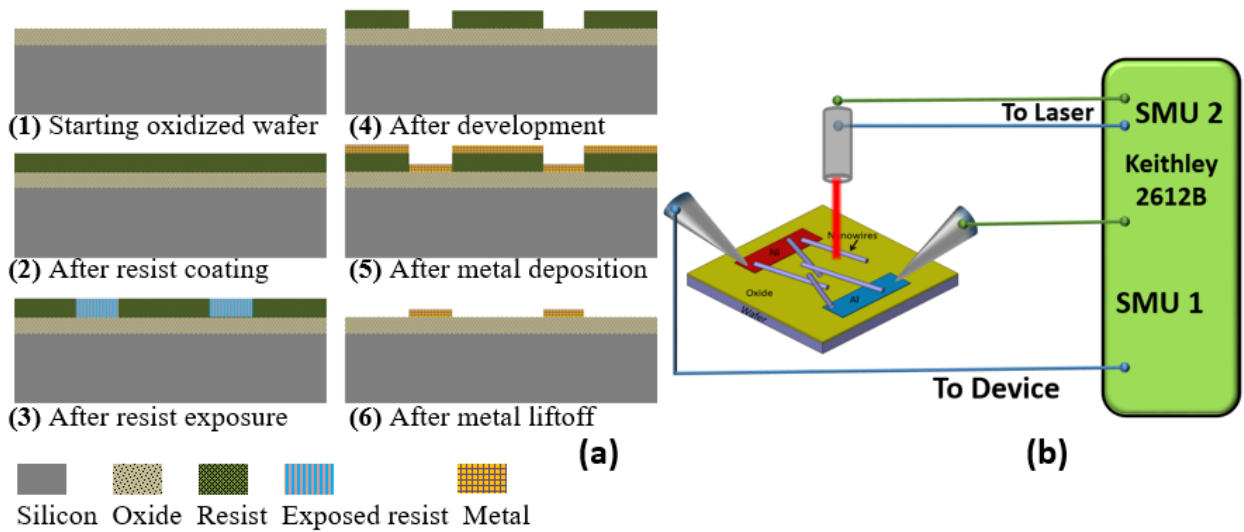

*Figure 1. (a) Fabrication steps for metal patterns on an insulating substrate, (b) I-V measurement setup.*

Figure 1(a) describes the fabrication steps of the metals-on-oxide test bench. Fabrication started with <100> Boron doped silicon wafer (resistivity of 1 to 5x10$^{-3}$ Ohm-cm) of 100 mm in diameter and 500 μm in thickness. The silicon wafer was wet oxidized to form a 1 μm SiO$_2$ layer on the top surface. The oxidized wafer was then cleaned with a nitrogen gas blower and primed with hexamethyldisilazane (HMDS) in a hot oven to enhance photoresist adhesion. Next AZ1512 photo-resist was spin coated at 3000 rpm for 45 seconds, which yielded a resist thickness of approximately $1.2 \mu m$. Immediately after coating, the wafers were baked at 100°C for 60 seconds on a hot plate. The resist-coated and baked wafers were exposed to UV light dose of 70 mJ/cm$^2$ with an AB-M aligner. The exposed resist was developed in MF319 developer for 30 seconds followed by removal of developed resist residue by discommending the wafer in a barrel Asher for 2 minutes with a power of 50 W and oxygen pressure of 1 Torr. Then the desired amount of adhesion and metals were deposited by using an electron-beam evaporator. Finally, the wafers were soaked in acetone to remove the unwanted metal along with the resist to obtain the desired metal patterns on the oxidized surface. The same steps were repeated for each metal patterned wafer used in this study. The first metal was Al (250 nm) with 15 nm Ti as an adhesion layer while the second metal was Ni (250 nm) with 15 nm Cr as an adhesion layer. Note that fabrication based on photo-lithography is good enough to pattern the electrode gaps ($10 \mu m - 100 \mu m$) used in this study; one may use e-beam lithography for even shorter gaps and better metal edge precision.

### c. Preparation of nanowire suspension and dispersion

For GeNW as well as SiNW, two types of suspensions, differing in NW density, were prepared by adding known weights of NWs, i.e., 15 mg and 5 mg, into a measured amount of toluene, 15 mL. Each of the mixtures was then sonicated for 5 minutes to obtain a uniform suspension of the NWs in toluene. Note that sonication also shortens the NWs, and therefore, for each sonication, a virgin suspension was used. Then, 2 μL of NW-in-toluene suspension was dispersed over the gap between Al and Ni contacts patterned on an oxidized Si surface, whose fabrication is described above. After the dispersion, the NWs settled out of solution via sedimentation and bridges the metal gap. The samples were kept as is for 24 hours to allow the toluene to evaporate completely. Although one can clearly see the NW density difference between the two cases in the SEM images in section III, the actual number of NWs between the metal pads in a device does not necessarily follow the precise ratio of NW amount to toluene.

Note that the proposed device fabrication process is relatively simple, and the test bench can also be reused after a quick wash-off of the NWs. Table 1 lists the samples that were studied.

*Table 1. Ge and Si NW samples studied in this work. The terms Dense and Sparse used in the text refer to the concentrations in this table, unless explicitely stated otherwise.*

| Sample Name | NW Amount (mg):Toluene (mL) |
|---|---|
| GeNW | 1:1 (Dense), 1:3 (Sparse) |
| SiNW | 1:1 (Dense), 1:3 (Sparse) |

### d. Measurement setup

The NW bridged metal-on-oxide devices were characterized both in dark and light conditions. Red laser diode was used as light source, with wavelength of 650 nm, power of 200mW, and spot size (diameter) of approximately 4 mm. Keithley source measure unit (SMU) 2612B was used to apply voltage across the gap between Al and Ni contacts, as well as to drive the laser diode. The same SMU was used to measure the currents. The setup shown in Figure 1(b) sat on an optical bench. Throughout the study, Al and Ni are employed as negative and positive electrodes, respectively. The Al-to-Ni distances (or device lengths) of 100 μm and 10 μm were considered. Because the laser spot size (~4 mm) was much larger than the device dimensions (100 μm and 10 μm), a uniform illumination over the entire RDNW device was attained. All the measurements were carried out in the ambient air at room temperature.

## III. Results and discussions
### a. Photo-response in RD-GeNWs

Figure 2 displays the mesurement results for dense RD-GeNWs with 100-μm electrode gap. The scanning electron micrograph (SEM) image of the device is in Figure 2(a). GeNWs tended to clump together upon dispersion. Figure 2(b) shows that the current-voltage (I-V) charecteristics are almost symmetric[31], and the photocurrents and dark currents increase non-linearly with bias. Such non-linearity is probably due to the work-function difference between the NW and metal contacts[32]. At a bias of 5 V, the photocurrent is approximately 85 nA compared to the dark current of approximately 40 nA.

More interestingly, Figure 2(c) shows the optoelectronic switching response of the RDNWs at zero bias and at a bias of 5 V. At zero bias, a significant photocurrent, ~40 pA, is observed. The zero-bias photocurrents in these highly disordered NWs indicate the effectiveness of using metal electrodes with different work functions. Notably, a mere random dispersion of NWs on dissimilar metal pads can extract such a high and distinct photocurrents at room temperature under visible-light illumination, even when no external bias is applied[32].

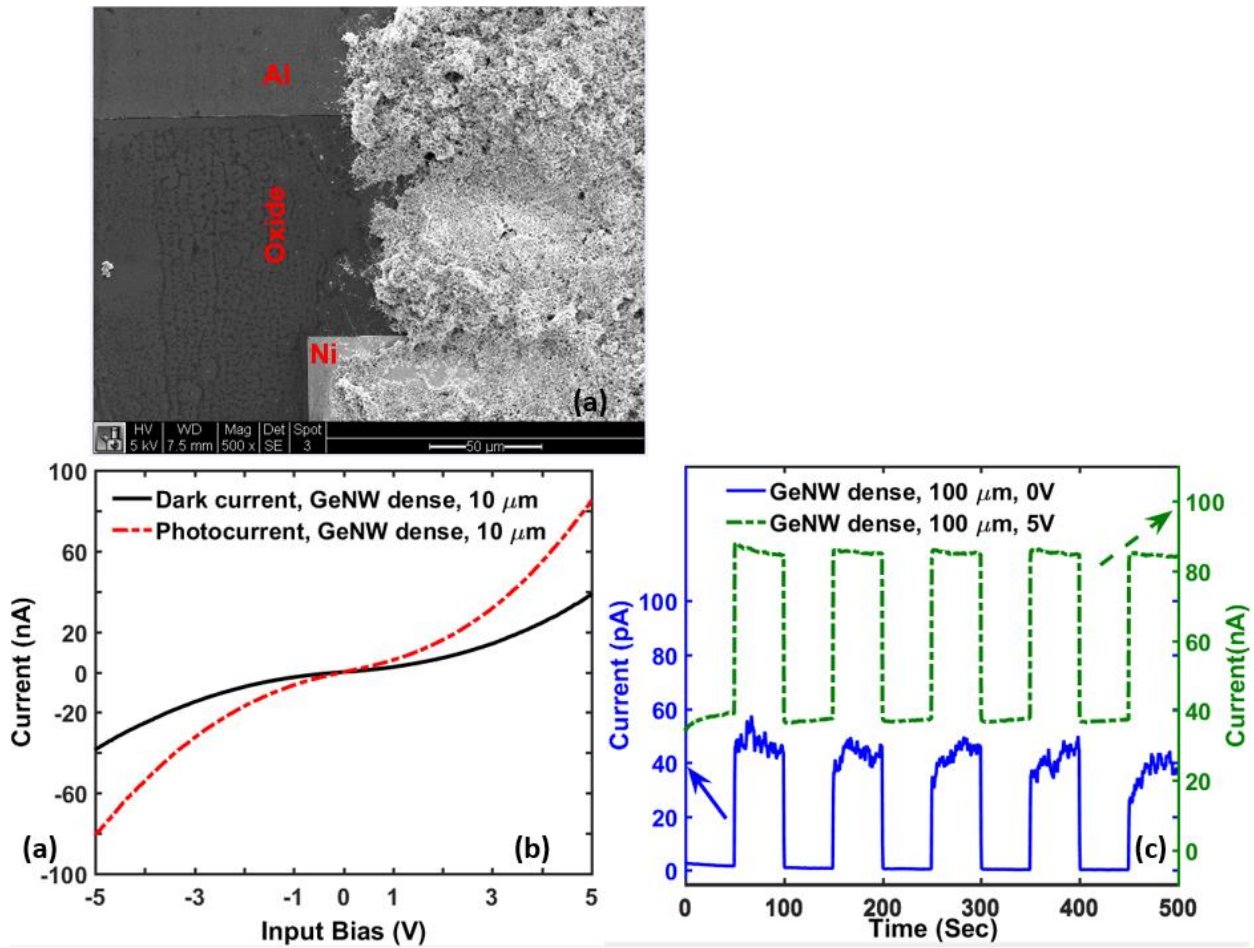

*Figure 2. (a) SEM image of dense RD-GeNWs with Al and Ni contacts (100 μm), (b) I-V characteristics in the dark (solid) and light (dashed), and (c) light switching at zero (blue solid, left y-axis) and 5 V (green dashed, right y-axis) bias in a dense RD-GeNWs device.*

The results from a sparse network of RD-GeNWs are shown in Figure 3. An SEM image of a sparse GeNW device with a gap of 100 μm between the Al and Ni pads is shown in Figure 3(a). The sparse GeNWs led to a decrease in the currents (Figure 3(b)) by ~2 orders of magnitude when compared to that of a denser GeNWs. The zero-bias photocurrents also decreased to approximately 1.5 pA. The reduction in current is apparently due to the relatively small number of active channels available for light absorption and for carrier transport. However, the switching response is still distinct and stable.

The NWs used in this study are typically several micrometers to tens of micrometers long, as noted in the synthesis section. Therefore, for a 100 µm electrode gap, no single NW can contact both electrodes directly (also evident from SEM images in Figures 2(a) and 3(a)). If, however, the gap is shorter than the NW length, it is likely that many individual NWs can directly contact both the electrodes, which can further improve the photoresponse. To test this proposition, the same NWs were deposited on a device with a 10-µm electrode gap setup (Figure 4(a) and (b)), and the optoelectronic response was measured. Figure 4(c) and (d) show the currents in both dense and sparse, 10-µm gap, RD-GeNW devices. Figure 4(c) shows the dark and photocurrents versus bias, while 4(d) plots the zero-bias photo-switching results. For the dense devices, the both photocurrent and the dark currents at 5 V (Figure 4(c)) increased by a factor of ~5 when compared to those of the 100-µm gap device (i.e, Figure 2(b)).

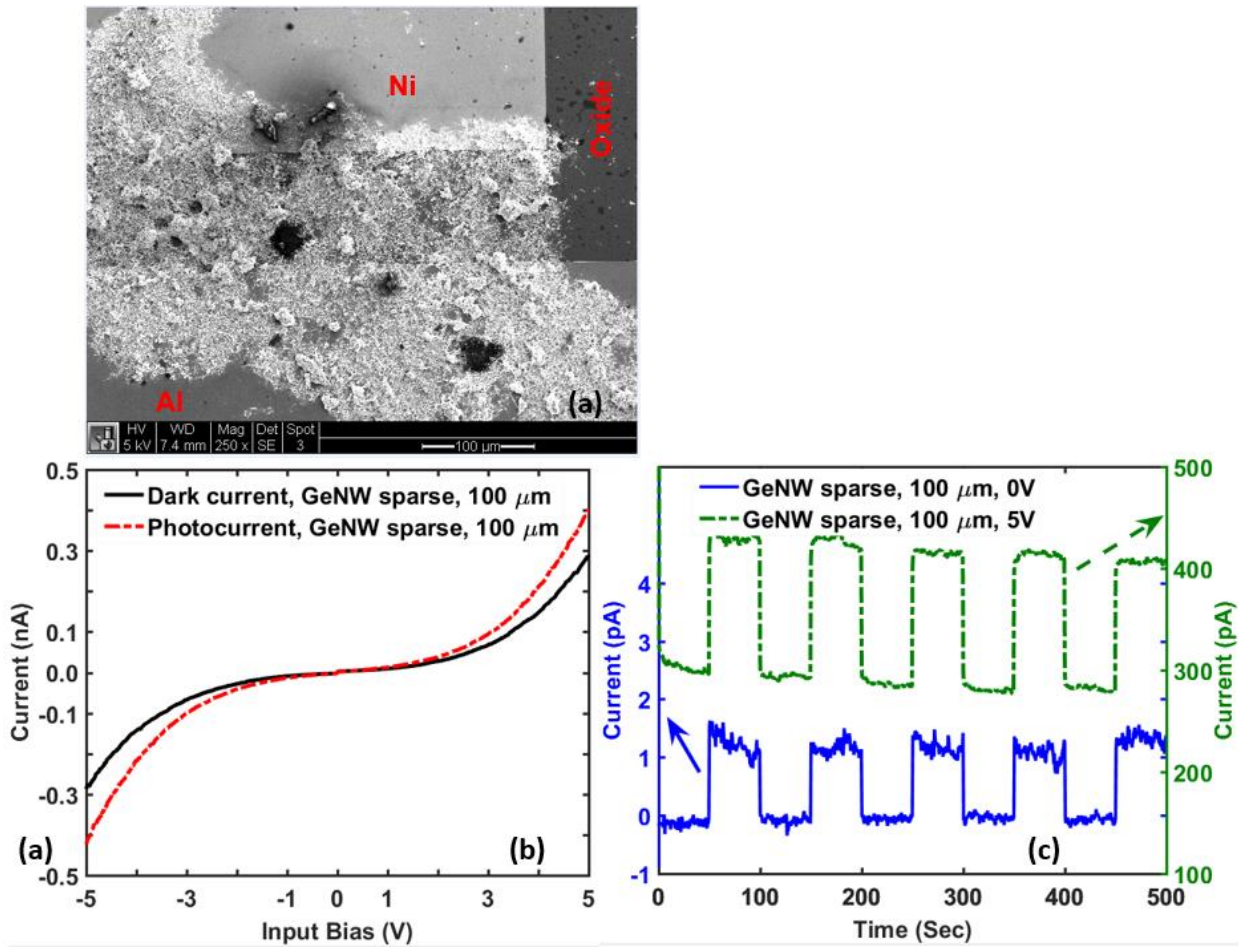

*Figure 3. (a) SEM image of sparse Ge-RDNWs with Al and Ni contacts (100 µm), (b) I-V characteristics in the dark (dashed) and light (solid), (c) Light switching in sparse Ge-RDNWs device at 5 V (green dashed, right y-axis) and zero bias (blue solid, left y-axis).*

However, the zero-bias photocurrent doubled. This finding confirms that devices with electrode gaps comparable to the length of the NWs more effectively produce a photocurrent than devices with a longer electrode gap. Similar channel-length dependence has been observed in carbon nanotube networks[33]. Although, due to the reduced area of the absorbing region, there are fewer photogenerated carriers available for the shorter device — the greater degree of direct, end-to-end contact between the NWs and both electrodes, as well as the stronger electric field that collects carriers more efficiently, more than compensated for this deficit in photogenerated charge carriers. Hence, greater absorption in long NWs might not always produce the expected higher photocurrent because carriers can only be collected from a distance on the order of the diffusion length[14]. Therefore, it can be inferred that, for the devices based on RD-NWs with asymmetric metal contacts, electrode-to-electrode distances smaller than the average NW length are more effective for photogenerated charge-carrier separation.

Moreover, an unexpected observation can be made when comparing the responses of the dense and sparse NW networks in Figure 4. It is often anticipated that the dense NW network has more channels to carry current and can also trap photon.

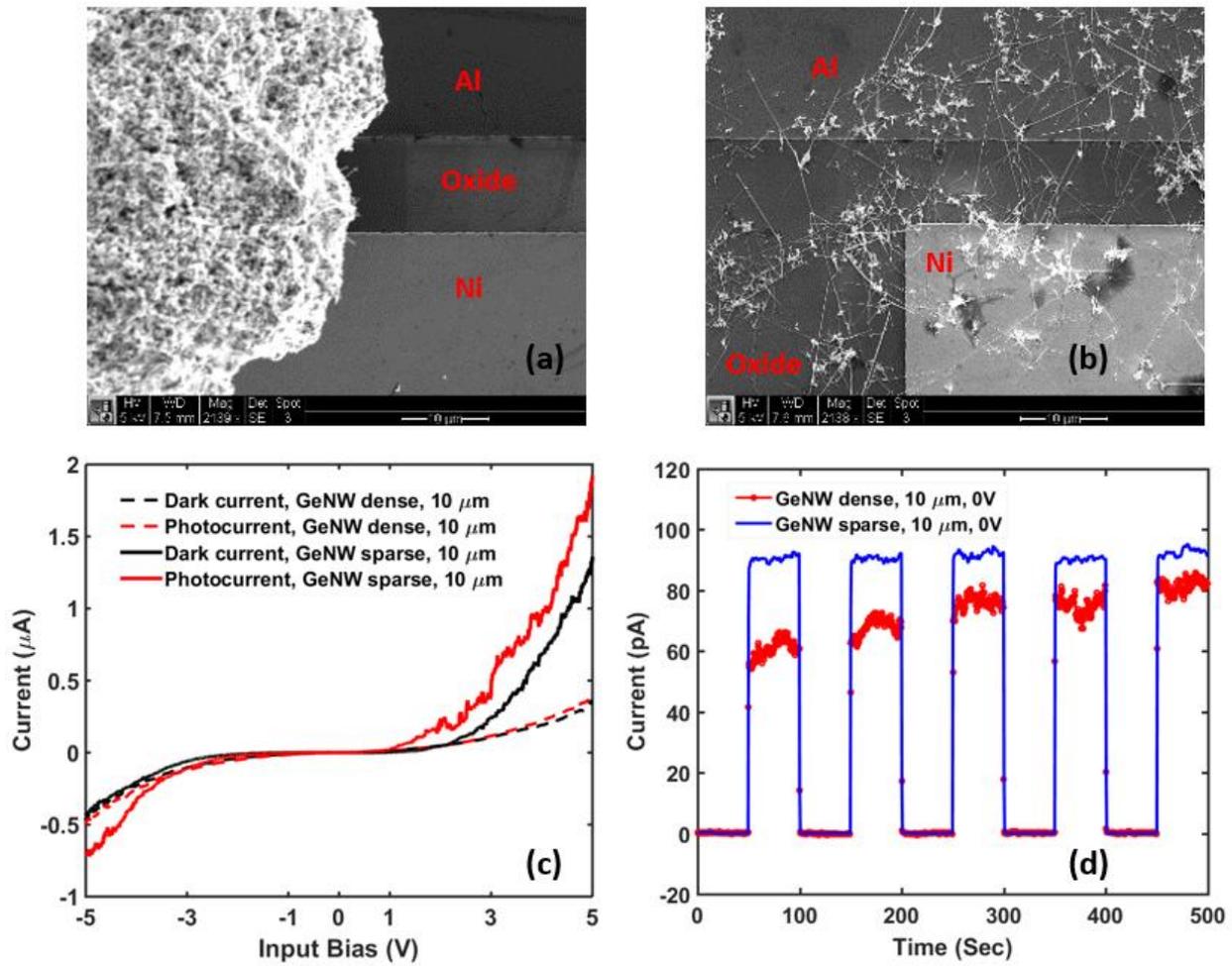

*Figure 4. SEM images of dense (a) and sparse (b) RD-GeNW devices with 10 μm metal gap, (c) I-V characteristics in dark and light, and (d) light switching at zero bias.*

more efficiently[7]. However, under a bias, the decrease in the number of dispersed NWs led to a significant increase in the current. The zero-bias photocurrent is also higher in the sparsely coated device (Figure 4(d), solid blue line).

To confirm that this particular result is generic, and not specific to GeNWs, SiNW network devices were characterized.

### b. Photo-Switching in RD-SiNWs

Figure 5 shows the results for SiNW devices with a 10 μm electrode gap. The SEM images of the dense and sparse SiNW network devices are shown in Figure 5 (a) and (b), respectively, while the dark and photo currents are plotted in Figure 5 (c) and (d). As in GeNW, the dense I-V characteristics are more symmetric than those of the sparse device. The SiNW networks also exhibit good photo-switching (Figure 5(e) and (f)). Here again, sparse NW devices also consistently produced more current than the dense devices; the photocurrent (at 5V bias) from the sparse NW network (~1.5nA) is almost an order greater than that of the dense (~200 pA) network. However, the current in the SiNW devciess is still lower than that in the similar GeNW devices (Figure 4(c)) because Ge has a greater conductivity as well as better optical absoprtion. The native Ge oxide is also less stable[34] than the native Si oxide and might allow for better NW-to-NW conductivity in GeNWs.

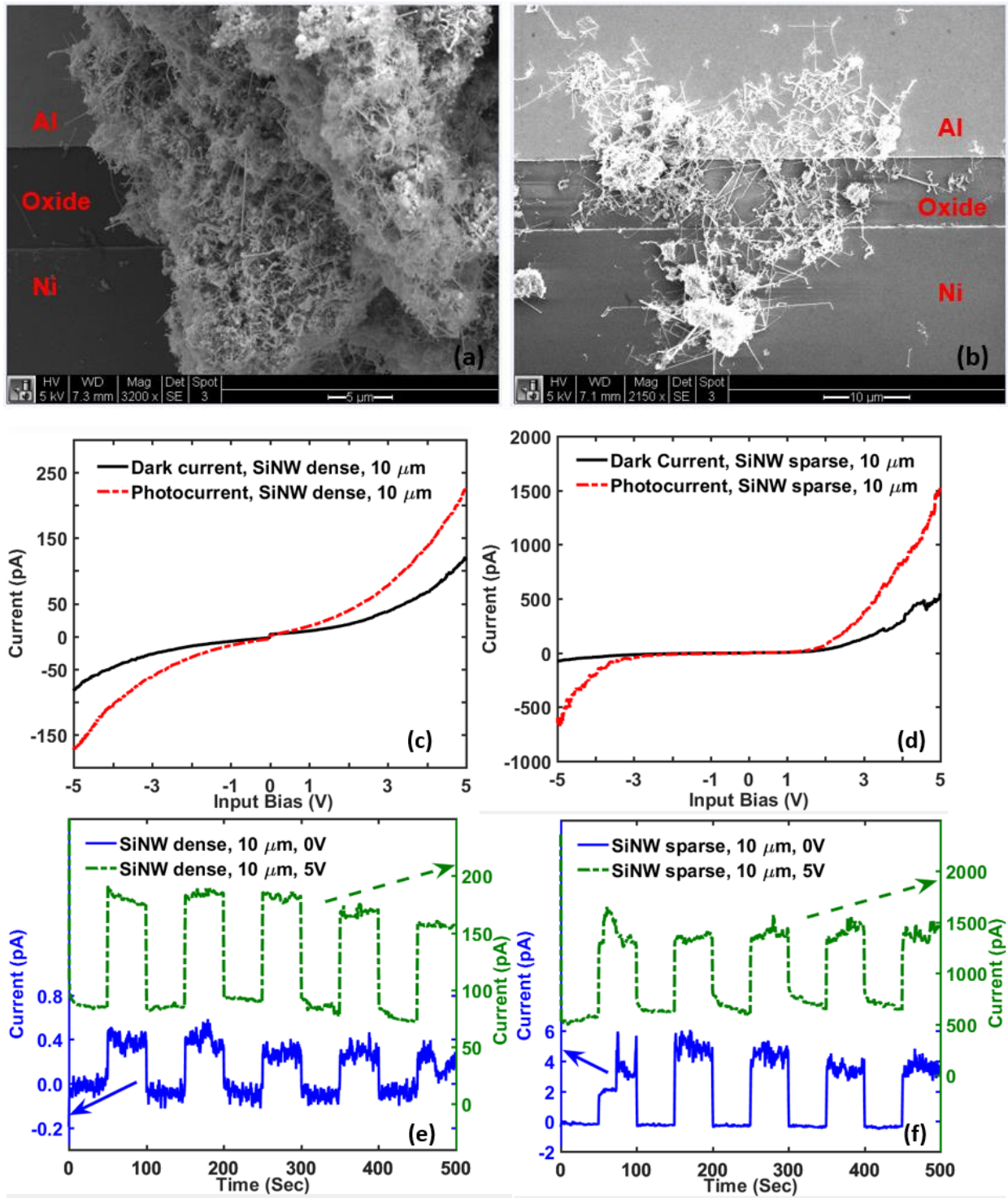

*Figure 5. SEM images of dense (a) and sparse (b) RD-SiNWs devices. I-V characteristics in the dark and light for both (c) dense and (d) sparse SiNW devices with a metal gap of 10 µm. The 5 V and zero-bias photo switching responses for the dense (e) and sparse (f) devices are also shown.*

### c. Discussions

To present some plausible explanations for the measured electrical characteristics, we take a bottom-up approach and consider the energy band diagrams for two different but simple current channels, shown in Figure 6. In the first case, shown in Figure 6(a), there is only a single NW between the two dissimilar work-function metals. The work function difference might induce band banding[14], but the exact degree of band bending will depend on the NW, the two metals and the two NW-metal contacts. Such bending can separate the photo-generated electrons and holes and generate a photocurrent. The amount of the current will also depend on the characteristics of the NW, the metal and the NW-metal contacts.

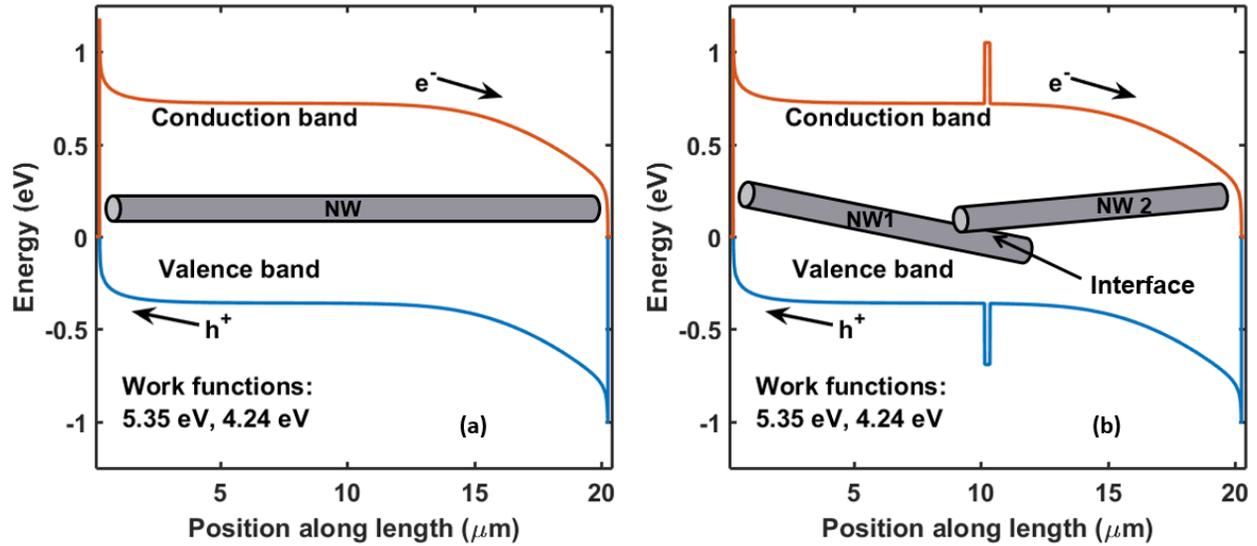

*Figure 6. Energy band diagrams of current path for a single NW (a), and multiple NWs (b).*

Figure 6(b) depicts the second case, in which a single current channel has two NWs instead of one. In addition to factors in the first case, *i.e.*, the NW, the metals and the NW-metal contacts, the NW-NW interface will have some effect on channel conductivity. The barriers at the middle of the energy bands represent non-ideal NW-NW interface due to presence of a nonconductive shell or native oxide at the NW surface or a surface depletion layer. Such a barrier can make the NW-NW interface the dominating factor and significantly reduce the channel conductivity. Obviously as the number of NWs involved in a single current channel increases, the number of NW-NW interfaces also increases and the current decrease might become more severe.

Depending on length and network density, dispersed NWs can be found in various possible orientations, and their conduction can be dominated by single or multiple NW channels.

- (i) If the NWs are sufficiently long and sparse, they can contact both electrodes directly, as shown in Figure 7(a). Therefore, the sparse newtork will effectively consist of many single-NW channels, such as the one in Figure 6(a) and lead to enhanced carrier conduction and greater current.
- (ii) The shorter NWs (i.e., shorter than the metal-to-metal gap) cannot individually contact both electrodes, and they can only interlink to bridge the electrode gap and form a device (cf. electrical percolation), as seen in Figure 7(b). In this case, the current channels are similar to the one in Figure 6(b). The carrier transport will be limited by the large NW-to-NW juction resistance due to the presence of a non-conducting shell or native oxide at the NW surface[30,35]. Thus, the higher the density of NWs, the better the interlink[33,36], and the larger the current.
- (iii) Irrespective of the length size, if the density of the NW network is higher than what is required for a one-NW-layer thick surface coverage, many NWs at the top might not contact either electrode. Rather, they will be in contact with the underlying NWs, which might be in direct contact with at least one of the electrodes (Al or Ni), as seen in Figure 7(c). Additionally most of the bottom NWs are likely in contact with only electrode at one end, while the other end of the NW is surrounded by other NWs, instead of contacting the other electrode. This scenario will again lead to current paths similar to the one in Figure 6(b). But, unlike case (ii), the current channels will involve more and more NWs as the network density increases, and the current will decrease due to increasingly inefficient NW-NW transport.

Therefore, in the dark, the sparse (and sufficientlylong) NWs produce more current due to better contact with the metal electrodes. When illuminated, the sparse NWs absorb photons directly and transport carriers effectively, thereby exhibiting higher photocurrents. However, in the denser NWs, the network is thick and most of the incident light is absorbed by the top NWs[37]. The photogenerated carriers inside these NWs need to be transported through other NWs to the electrode, and therefore, the NW-NW junction resistance limits the current flow, leading to smaller current values in dense NW devices. This phenomenon is a result of inefficient nanoscale transport.

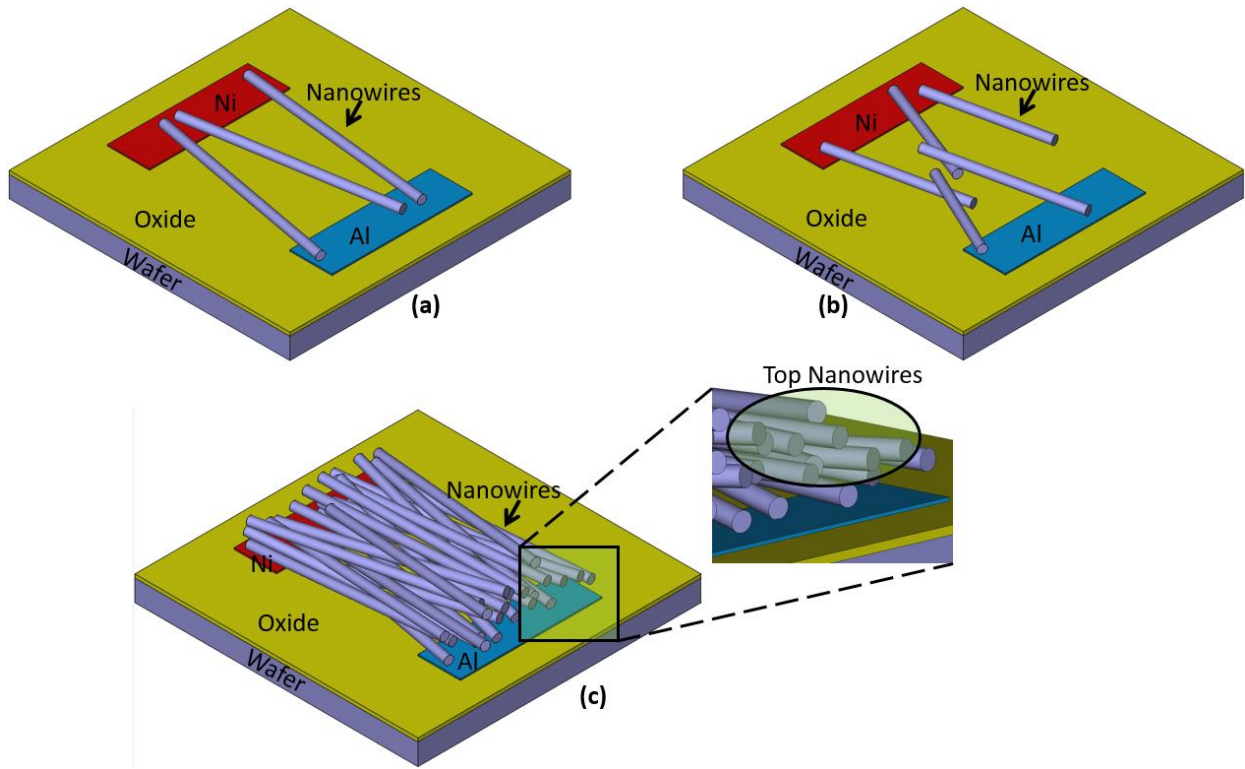

*Figure 7. RDNW device schematics. (a) long sparse NW network (inset shows bandstructure of a Ni-NW-Al channel), (b) short, sparse NW network (inset shows bandstructure of a Ni-NW1-Oxide-NW2-Al channel), (c) long, dense NW network, inset shows zoom in on one end of some of the top NWs.*

IV. **Conclusions**

In summary, Ge- and SiNW-based low-cost randomly dispersed networks were fabricated and their potential for self-powered photo-switching has been demonstrated. These devices can efficiently trap and absorb the incident light and collect the photogenerated excitons, as evidenced by photocurrent measurements. Larger currents can be extracted from GeNWs than from SiNWs. GeNW devices exhibit better photo-switching because of greater conductivity and inter-NW transport. The 'anomalous' increase in current as the density of the randomly dispersed NW networks decreases, when the electrode gap is comparable to the NW lengths, is a general phenomenon of the devices tested and has been clearly explained. The consistency of the results indicates that this anomaly is applicable to nanoscale devices with electrodes of either similar or dissimilar work function and that there is an optimum number of dispersed NWs for obtaining the best optoelectronic response.

**Acknowledgements** We acknowledge the use of the UW Molecular Analysis Facility (MAF) and Microelectronics Laboratory for the SEM and electrical characterization, and the UW NNIN Washington Nanofabrication Facility (WNF) for the device fabrication.

The work of M. Golam Rabbani and M. P. Anantram was supported by the National Science Foundation under Grant No. 1001174. M. P. Anantram were also partially supported by the QNRF grant (NPRP 5 – 968 – 2 – 403) and the University of Washington. Amit Verma, Mahmoud M. Khader, and Reza Nekovei were supported by the QNRF grant (NPRP 5 – 968 – 2 – 403). Brian A. Korgel acknowledges funding of this work by the Robert A. Welch Foundation (Grant no. F-1464) and the National Science Foundation (CHE-1308813). Sunil R. Patil was supported by UGC-India Grant No. F.5-50/2014(IC).